\documentclass[times]{article}
\usepackage{isita2004}
\usepackage{graphicx}
\title{An Efficient MN-Algorithm for Joint Source-Channel Coding}
\name{Haggai KFIR$^{\dagger}$, Eyal SHPILMAN$^{\dagger}$ and Ido KANTER$^{\dagger}$%
\thanks{This work was supported by the Israel Academy of Science .}
}

\address{
\begin{tabular}{c}
$^{\dagger}$ Minerva Center and Department of Physics, Bar-Ilan
University Ramat-Gan, 52900 Israel\\
E-mail: kfirha1@mail.biu.ac.il, shpilme@mail.biu.ac.il,
kanter@mail.biu.ac.il
\end{tabular}
}

\begin{document}

\maketitle
\sloppy
\begin{abstract}
Belief Propagation (BP) decoding of LDPC codes is extended to the
case of Joint Source-Channel coding. The uncompressed source is
treated as a Markov process, characterized by a transition matrix,
$T$, which is utilized as side information for the Joint scheme.
The method is based on the ability to calculate a Dynamical Block
Prior (DBP), for each decoded symbol separately, and re-estimate
this prior after every iteration of the BP decoder. We demonstrate
the implementation of this method using MacKay and Neel's LDPC
algorithm over $GF(q)$, and present simulation results indicating
that the proposed scheme is comparable with Separation scheme,
even when advanced compression algorithms (such as AC, PPM) are
used. The extension to 2D (and higher) arrays of symbols is
straight-forward. The possibility of using the proposed scheme
without side information is briefly sketched.
\end{abstract}

\section{INTRODUCTION}

The Shannon separation theorem \cite{Shannon-48,Cover}, states
that source coding and channel coding can be performed separately
and sequentially, while maintaining optimality. However, this is
true only in the case of asymptotically long blocks of data. Thus,
considerable interest has developed in various schemes of joint
source-channel coding, where the inherent redundancy of the source
is utilized for error correction, possibly with the aid of some
side information (see, for instance, \cite{shamail6}). Combining
the two processes may be motivated by reducing the total
complexity of the procedure, and by some gain in the overall
performance. Moreover, some uncompressed files (e.g. bitmap, text)
are expected to be resilient to single bit errors, which may
corrupt entire blocks in the case of the Separation scheme.

Shannon's lower bound for the channel capacity of a binary
symmetric channel (BSC) with flip probability $f$, bit error rate
$p_b$ and source entropy $H(src)$ per bit is given by
\cite{Shannon-48}:
\begin{equation}\label{bsc_capacity}
    C=\frac{1-H_{2}\left(f\right)}{H\left(src\right)-H_2(p_b)},
\end{equation}

\noindent where
$H_2\left(x\right)=-xlog_2\left(x\right)-\left(1-x\right)log_2\left(1-x\right)$,
is the entropy of $x$, and the capacity, $C$, is the maximal ratio
between the source length $k$ and the transmitted length $m$.

In this paper we propose an extension of the
Low-Density-Parity-Check codes (LDPC) \cite{Gallager} decodeing
algorithm, primarily designed for i.i.d. sequences, to the case of
uncompressed data. Our approach is to regard the source sequence,
$\{s_n\}$, as driven from some memoryless stationary Markov
process with a finite alphabet $s_n\in{\{0,1,2...q-1\}}$, and
transition matrix $T$ of dimensions $q\times q$, that describes
the probability of transition from symbol $i$ to symbol $j$:
$t_{ij}=P(s_{n+1}=j\mid s_n=i)$. The Markov Entropy (per symbol)
of such a process is given by:
\begin{equation}\label{Markov_entropy}
    H=-\sum^{q}_{i=1}{P(i)}\sum^{q}_{j=1}{P(j\mid i)log_2[P(j\mid
    i)]},
\end{equation}

\noindent where $P(i)$  is the stationary solution of the Markov
process. The entropy \emph{per bit}, $H/\log_2q$, ($\log_2q$ being
the number of bits in the binary representation of a symbol) can
be utilized as $H(src)$ in Eq. (\ref{bsc_capacity}).


 Neighboring symbols in a Markov sequence are
\emph{correlated}. Hence, information about symbols $s_{n-1}$ and
$s_{n+1}$, immediately implies some knowledge about $s_n$, too.
The main contribution of this work, is a method of incorporating
this additional knowledge into the Belief Propagation decoding
scheme.



\section{MN ALGORITHM}

Our joint source-channel scheme is based on Mackay and Neel's
algorithm (a thorough introduction may be found in \cite{MNlong}),
a variant of the earlier Gallager code \cite{Gallager}. Although
originally proposed for the binary field, extending the MN
algorithm to higher finite fields is straight-forward as
demonstrated in \cite{MNgfq}. The original motivation for moving
to higher fields was reducing the number of edges (and short
loops) in the code's graph. For our purpose, this enables us to
treat Markov sequences with a richer alphabet, consisting of
$q=2^i$ symbols ($i$ being an integer). The algorithm consists of
two sparse matrices known both to the sender and the receiver: $A
(m\times k)$, and $B (m\times m)$, where $k$ is the source block
length, $m$ is the transmitted block length, and the code rate
being $R=k/m$. All non-zero elements in $A$ and $B$ are from
$\{1,2...q-1\}$, and $B$ must be invertible. Encoding of a source
vector $s$ into a codeword $t$ is performed (all operations are
done over $GF(q)$) by:
\begin{equation}\label{MNencode}
    t=B^{-1}\cdot A\cdot s.
\end{equation}
$t$ is converted to binary representation and transmitted over the
channel. During transmission, noise $n$ is added to $t$, therefore
the received vector is $r=t+n$. Upon receipt, the decoder
reconverts $r$ back to the original field, and computes the
syndrome vector $z=B\cdot r$. The receiver then faces the
following decoding problem:
\begin{equation}\label{MNdecode}
    z=B\cdot(t+n)=B\cdot(B^{-1}\cdot A\cdot
    s+n)=[AB]\cdot x,
\end{equation}

\begin{figure}
\centering
\includegraphics[width=2.0in]{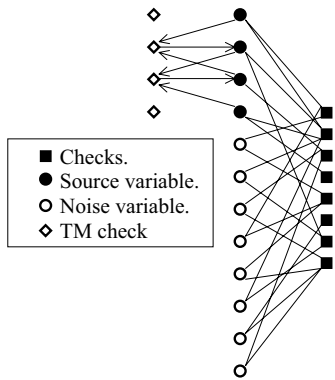}
\caption{The Dynamic Block Prior visualized as a third layer
attached to the bipartite graph of the LDPC code.} \label{3layer}
\end{figure}

\noindent where square brackets denote appending of matrices, and
$x$ is a concatenation of $s$ and $n$. The decoding problem can be
visualized as a bipartite graph (Fig \ref{3layer}), the elements
of $x$ (circles) and $z$ (squares) are termed "variable" and
"check" nodes, respectivly. The edges of the graph correspond to
the nonzero elements in $[AB]$. For the MN algorithm, one should
further distinguish between "source variables" - the $s$ elements
in $x$ (filled circles), and "noise variables" - the $n$ elements
in $x$ (empty circles). The decoding problem may be solved using
the Belief Propagation (BP) (or sum - product) algorithm
\cite{MNlong,MNgfq}. BP is an iterative algorithm with two
alternating steps, horizontal pass (check$\rightarrow$ variable
messages) and vertical pass (variable $\rightarrow$ check
messages). During the vertical pass, some prior knowledge is
assigned to each decoded symbol, according to the assumed
statistics (for the i.i.d. case this would simply be:
$Pr(s=j)=1/q$ for all the source symbols). The key point here is
that one can re-estimate and re-assign these priors after every
iteration \emph{individually for each decoded symbol}
 \cite{IdoHanan}. The outcome of each iteration is an a-posteriori
probability $Q^a_i=Pr(x_i=a)$, for each symbol (both source and
noise). The MN decoder is linear in the size of the source block,
$k$, with complexity $O(kqu)$ (per iteration), where $u$ is the
average number of checks per symbol \cite{MNcomplx1,MNcomplx2}.

\begin{figure}
\centering
\includegraphics[width=3.0in]{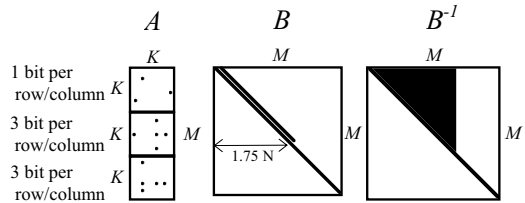}
\vspace{-1 cm} \caption{Kanter-Saad \cite{KSbsc} construction of
the matrices $A$, $B$, $B^{-1}$ for rate $1/3$, black regions
denote nonzero elements. } \label{KS_construction}
\end{figure}

A proper construction of the matrices $A$ and $B$ is crucial in
order to ensure nearly capacity-achieving performance. In this
work we follow the Kanter and Saad (KS) constructions
\cite{KSbsc,KS-Gaussian}, which are very sparse, simple to
construct, and preform very close to the bound. The $B$ matrix has
a systematic construction: diagonal and sub diagonal, which
simplifies computation tasks \cite{MNcomplx1,KS-Gaussian,saad1}.
The KS construction for $R=1/3$, $GF(2)$, is schematically
displayed in Fig. \ref{KS_construction}, black regions denote
nonzero elements. Extending the construction to higher $GF(q)$ is
done by randomly replacing the nonzero elements with elements of
the corresponding field. Although constructed originally for
i.i.d. sources, we successfully used KS matrices for uncompressed
sources, however, we mention the possibility of improving the
performance by devising better codes.

The MN algorithm is also applicable for the case of an Additive
White Gaussian Noise (AWGN) channel \cite{KS-Gaussian}. The binary
transmitted vector (assumed for simplicity to be $\pm1$) is
corrupted by noise with zero mean and variance $\sigma^2$, hence,
the received vector, $r^R$, is real valued. The \emph{binary}
received vector, $r$, is determined by hard-decision, namely,
$r_i=+1$ if $r^R_i>0$. The probability of the transmitted bit
$t_i=\pm 1$ is given by:
\begin{eqnarray}\label{gaussian_t_prob}
    P(t_i=\pm 1|r^R_i)&=&\frac{e^{-(t_i-r^R_i)^2/2\sigma^2}}
    {e^{-(t_i-r^R_i)^2/2\sigma^2}+e^{-(t_i+r^R_i)^2/2\sigma^2}} \nonumber\\
    &=&\frac{1}{1+e^{-2t_ir^R_i/\sigma^2}},
\end{eqnarray}
\noindent and the probability of an error in the $i^{th}$
hard-decision bit is given by:
\begin{eqnarray}\label{gaussian_n_prob}
    P(n_i=1|r^R_i)=\frac{1}{1+e^{2|r^R_i|/\sigma^2}}.
\end{eqnarray}
Eq. (\ref{gaussian_n_prob}) is used for calculating a prior for
each noise variable. Apart of these modifications, the MN
algorithm for an AWGN channel is identical to the BSC case.

The channel capacity for AWGN is given by \cite{Cover}:

\begin{equation}\label{capacity_AWGN}
    C=\frac{1}{2\cdot H(src)}\log(1+\frac{1}{\sigma^2})
\end{equation}

For binary source messages, (rather then real source messages),
however, there exist a tighter bound, \cite{KS-Gaussian}:
\begin{eqnarray}\label{binary_capacity}
    C=\frac{1}{H(src)}(-\int dyP(y)\log P(y)+ \nonumber\\ \int {dy P(y|x=x_0)\log P(y|x=x_0)})
\end{eqnarray}
where x is the transmitted bit, $x_0=\pm 1$ and $y$ is the
received (corrupted) bit, with
\begin{equation}\label{P_y}
    P(y)=\frac{1}{2\sqrt{2\pi\sigma^2}}\left[e^{-(y-x)^2/2\sigma^2}+e^{-(y+x)^2/2\sigma^2}\right].
\end{equation}

\section{DERIVING THE DYNAMICAL BLOCK PRIORS}
In every iteration of the MN algorithm, a better estimate of each
variable node is attained (on average). In this section we shall
describe our method of incorporating the statistical knowledge
about the source, and these local estimates. Consider three
successive symbols $s_{n-1},s_n,s_{n+1}$ in a sequence generated
by a Markov process with transition matrix $T$ and alphabet
$GF(q)$. The probability of a triplet $a,b,c$ is given by
\cite{Measur}:

\begin{eqnarray}\label{Pabc}
  P(a,b,c)&=&P(a,b)\cdot P(c|a,b)=P(a,b)\cdot P(c|b)\nonumber\\
&=&\frac{P(a,b)P(b,c)}{P(b)}
\end{eqnarray}

\noindent where use has been made of the Bayes Rule:
$P(x,y)=P(x)\cdot P(y|x)$, and the fact that the process is
memoryless. Now, given the a-posteriori probabilities for the
first and last symbols in the triplet: $Q_{n-1}^a=Pr(s_{n-1}=a)$
and $Q_{n+1}^c=Pr(s_{n+1}=c)$, one can calculate a prior for the
probability that $s_n=b$:


\begin{eqnarray}\label{Prior(b)}
  Pr(s_n=b)=\frac{1}{Z}\cdot \sum_{a,c=1}^q{P(a,b,c)\cdot Q_{n-1}^a}\cdot Q_{n+1}^c=\nonumber \\
  =\frac{1}{Z}P(b)^{-1} \!\!\left(\sum_{a=1}^q{P(a,b) Q_{n-1}^a}\!\!\right)
 \left(\sum_{c=1}^q{P(b,c) Q_{n+1}^c}\!\!\right),
\end{eqnarray}

\noindent where $Z$ is a normalization constant such that:
$\sum_{b=1}^q{Pr(s_n=b)}=1$. We term Eq. (\ref{Prior(b)}) the
Dynamical Block Prior (DBP).

The extension of the MN algorithm to the joint source-channel case
consists of the following steps:\begin{enumerate}
    \item A binary sequence of $k\cdot log_2(q)$ bits is converted to $k$ $GF(q)$ symbols.
    \item The encoder measures $T$ and $P(a)$ for all
    the $q$ symbols over the source, and transmits reliably this side
    information to the decoder.
    \item The source is encoded according to
    (\ref{MNencode}), then reconverted to binary representation
    and transmitted over the BSC.
    \item The decoder maps the received signal back to GF($q$), and
    performs the regular decoding
    (\ref{MNdecode}), but after every iteration of the BP, the prior for each source symbol is
    recalculated according to (\ref{Prior(b)}).
\end{enumerate}
The complexity of calculating the $q$ priors for a single symbol
according to the posteriors of its neighbors is reduced from $q^3$
in the naive calculation, to $q^2$ by Eq. (\ref{Prior(b)}). The
decoder's complexity remains linear, with total complexity of
$O(kqu+kq^2)$ per iteration. The above-mentioned procedure may be
thought of as adding a layer to the bipartite random graph
represented by the matrix $[AB]$: The DBP's, Eq. (\ref{Prior(b)}),
are messages passed only among \emph{source} variable nodes, which
are spatially related. In Fig. \ref{3layer}, the diamonds
represent this new (directional) layer, which connects neighboring
source nodes. We note that the possibility of extending this
scheme to Gallager codes is an open question, since the source is
not explicitly represented in the graph.

\section{SIMULATION RESULTS}
We report here results for a BSC with rate $R=1/3$,  and for an
AWGN with rate $1/4$, using the corresponding KS constructions for
$A$ and $B$ devised in \cite{KSbsc,KS-Gaussian}. Other rates,
constructions and block length were also checked. Random vectors
of length $L=10^4$ bits ($9,999$ for $q=8$) were generated by the
Markov process, then mapped to a vector in $GF(q)$ with length
$k=L/log_2[q]$, and were encoded and decoded as described in the
previous section. For each reported result, at least 1000 sample
vectors were generated and transmitted.

\subsection{Estimating The Code's Threshold}

The threshold for infinite source length, $k \rightarrow \infty $,
is estimated from the scaling argument of the convergence time,
which was previously observed for $q=2$ \cite{KSbsc,KS-Gaussian}.
The convergence time, measured in iterations of the MN algorithm,
is assumed to diverge as the level of noise approaches the
threshold from below. More precisely, we found that the scaling
for the divergence of $t_{med}$ is {\it independent of $q$} and is
consistent with:
\begin{equation}
t_{med}(f) \propto {1 \over f_\infty-f}~~~~;~~~~ t_{med}(\sigma)
\propto {1 \over \sigma_\infty-\sigma}\label{scaling}
\end{equation}
for a BSC, and an AWGN channel, respectivly.

This extrapolation is independent of $k$ \cite{SM_of_JSC} (for
$k>>1$), so by monitoring $t_{med}$, for moderate $k$, the
threshold can be found by a linear fit. (see the inset of Fig.
\ref{pb_vs_f}) Note that the estimation of $t_{med}$ is a simple
computational task in comparison with the estimation of low bit
error probabilities for large $k$, especially close to the
threshold. We also note that the analysis is based on $t_{med}$
instead of the \emph{average} number of iterations, since we wish
to prevent the dramatic effect of a small fraction of samples with
slow convergence or no convergence.

\subsection{BSC Simulations}
 Some selected results are presented in Table
\ref{res_table}. The columns correspond to: the field size $q$;
the source length in symbols, $k$; the entropy (per bit) of the
source $H$ \cite{Tgf4}; and the corresponding maximal noise
$f_{Sh}$ (Eq. (\ref{bsc_capacity})); the critical noise, $f_c$, up
to which the bit error rate $p_b\leq10^{-5}$; and the threshold,
$f_\infty$, Eq. (\ref{scaling}).

\begin{table}
\renewcommand{\arraystretch}{1.3}

\begin{center}
\caption{Simulation results for BSC with rate $1/3$.}
\label{res_table}

\begin{tabular}{|c|c|c|c|c|c|}
  \hline
  $q$ & $k$ & $H$& $f_{Sh}$ & $f_c$ & $f_\infty$ \\
  \hline
  $4$&$5000$&$0.49$&$0.266$&$0.215$&$0.244$ \\
  $8$ & $3333$ & $0.471$ & $0.271$ & $0.223$&$0.243$\\
  $16$ & $2500$ & $0.49$ & $0.266$ & $0.21$  &$0.236$\\
  \hline
\end{tabular}

\end{center}

\end{table}
In order to compare the joint and the Separation schemes, the
generated samples were concatenated to strings of size
$L=10^5-10^6$ bits, and compressed using two advanced compression
algorithms: Prediction by Partial Match (PPM) \cite{PPM}, and
Arithmetic Coder (AC) \cite{AC}. In Table \ref{compar2sep}, we
preset the compression ratio of the source vectors using each
method ($\%AC,~~\%PPM$); the ratio between the compressed
sequences and the transmitted blocks, $m$, (for the comparison, we
use the same transmitted size as for the Joint scheme, $m=3k$);
and the maximal noise level for this new rate for i.i.d. source,
$f_{AC}, f_{PPM}$. Since we assume an \emph{optimal} decoder,
these noise levels should be compared to $f_\infty$ for the joint
scheme. In all cases, the threshold of the proposed Joint scheme
for $k\rightarrow \infty$ is comparable with the Separation
scheme. One should recall that these results may be improved by
advanced codes.
\begin{figure}
{\centering
{\includegraphics[width=2.5in]{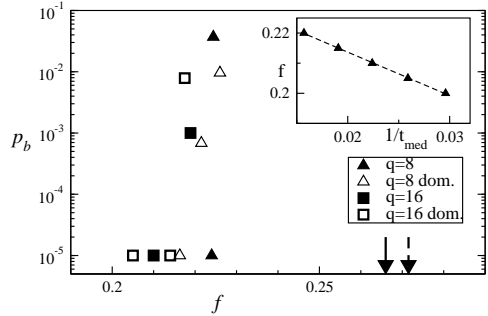}}
\par}
\caption{\label{pb_vs_f}Bit error rate $p_b$ vs noise level $f$,
triangles for $q=8$, squares for $q=4$, filled symbols: use of all
elements of $T$ as side information, empty symbols: use of only
$q$ dominant elements of $T$. Inset: scaling behavior for median
convergence time $t_{med}$. $f_\infty=0.242$, is found by linear
fit.}
\end{figure}
In Fig. \ref{pb_vs_f}, $p_b$ is plotted against the noise level of
the channel, $f$, for the examples of Table \ref{res_table}.
Filled triangles represent $q=8$; filled squares represent $q=16$;
the empty symbols in this figure refer to an approximation that
will be described in the following section. The dashed (full)
arrow marks $f_{Sh}$ for $q=8~  (16)$. The inset of Fig.
\ref{pb_vs_f} demonstrates the extrapolation of $f_{\infty}$ from
the convergence time Eq. (\ref{scaling}): for $q=8$, $f(t_{med})$
is plotted against $1/t_{med}$, $f_{\infty}$ is then recovered by
a linear fit.
\begin{table}
\renewcommand{\arraystretch}{1.3}

\begin{center}
\caption{Critical noise level for Separation scheme using
Arithmetic Coder and Prediction by Partial Match compression
algorithms. $f_{AC}$ and $f_{PPM}$ should be compared to
$f_\infty$ in Table \ref{res_table}.}
 \label{compar2sep}
\begin{tabular}{|c|c|c|c|c|c|c|}
  \hline
  $q$ & $\%AC$ & $R_{AC}$ & $f_{AC}$& $\%PPM$ & $R_{PPM}$ & $f_{PPM}$ \\
  \hline
  $4$&$58.4\%$&$0.195$&$0.247$&$56.6\%$&$0.189$ &$0.25$  \\
  $8$ & $58.1\%$ & $0.194$ & $0.248$&$55.5\%$& $0.185$ & $0.253$ \\
  $16$ & $60.5\%$ & $0.201$ & $0.243$  &$59.4\%$&$0.198$&$0.245$\\
  \hline
\end{tabular}

\end{center}
\end{table}
\subsection{AWGN Simulations}
\begin{figure}
{\centering \resizebox*{0.4\textwidth}{0.25\textheight}
{\includegraphics[width=2.5in,angle=270]{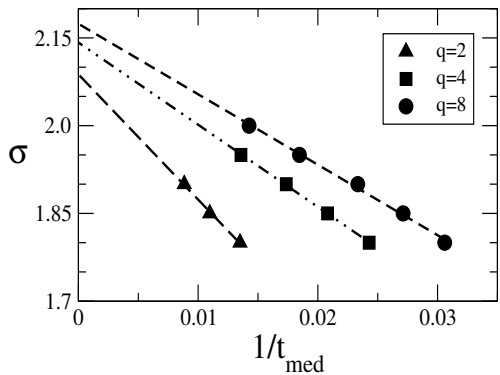}}
\par}
\caption{Scaling behavior for AWGN channel: $\sigma$ is plotted
vs. $1/t_{med}$ for various field sizes, $q$.}
\label{Gaussian_res}
\end{figure}

Binary sequences of length $k=10^4$ were generated using the
following transition matrix:
\begin{equation}
\label{T_c1_078} T= \left(
\begin{array}{cc}
  0.89 & 0.11 \\
  0.11 & 0.89 \\
\end{array}
\right),
\end{equation}
having Markov Entropy $H(src)=0.5$. For rate $R=1/4$, this entropy
corresponds to maximal noise, Eq. (\ref{binary_capacity}),
$\sigma_{Sh}=2.298$, ($-4.2$ Db). The sequences were transmitted
over an AWGN channel, using three different fields:
$GF(2),~GF(4),~GF(8)$. Fig. \ref{Gaussian_res} presents the
scaling behavior, Eq.(\ref{scaling}), for these fields (triangles,
squares, and circles, respectively). The symbols mark working
points with $p_b\leq 10^{-5}$, and were used for estimating the
corresponding thresholds:
$\sigma_\infty(q=2)=2.08,~\sigma_\infty(q=4)=2.14,~\sigma_\infty(q=8)=2.17$.
It is evident that as $q$ increases, both $\sigma_c(q)$ and
$\sigma_{\infty}(q)$ improve.
\section{REDUCING THE AMOUNT OF SIDE INFORMATION}

The results of the previous sections indicate that the performance
of the presented joint coding is not too far from Shannon's lower
bound and, most probably, using an optimized code, the channel
capacity can be nearly saturated. However, for a finite block
length, the main drawback of our algorithm is the overhead of the
header (i.e. the transmitted side information) which must be
encoded and transmitted reliably. One has to remember that the
size of the header, ($T$), scales with $q^2$ where the precision
of each element is of the order $O(\log k)$. This overhead is
especially intolerable in the limit where: $q^2 \log (k)/k \sim
O(1)$.
Note that this is indeed the situation even for very large
messages, $k=10^6$, and a symbol size of $8$ bits (a "char",
$q=256$).

This point may be tackled by observing that for a process with low
entropy, characterized by enhanced repetitions and correlations,
$T$ is dominated by a small number of elements, while the rest of
the elements are negligible. We therefore repeated our
simulations, using only the $q$ largest elements in $T$ as side
information. The decoder would then set all other elements in each
row of $T$ equally, to obey the normalization condition
$\sum_jT_{ij}=1$. In Fig. \ref{pb_vs_f} the empty
squares/triangles represent working points for the algorithm with
$q=8/16$. In both cases, the critical noise level $f_c$ is only
slightly decreased, but the size of the side information becomes
considerably smaller.


\section{JOINT SOURCE-CHANNEL CODING WITH THE LACK OF SIDE INFORMATION}

%

In this section we describe how the Markovian decoder can be
implemented without any transmission of side information. The key
points are the special properties of the KS construction (Fig.
\ref{KS_construction}): the first $k$ rows of $A$ are
characterized by one non-zero element per row and column, where
the first $k$ rows of $B$ are characterized by $2$ non-zero
elements. Furthermore, due to the systematic form of $B$, each row
cannot be written as a linear combination of the other rows.
Hence, the first $k$ bits of the syndrome vector $z$, are equal
(up to a simple permutation) to the source, with an effective flip
rate, $f_{\emph{eff}}$ . For $GF(2)$ for instance,
$z_j=s_i+n_j+n_{j+1}$ ($i$ marks the position of the nonzero
element in the $j^{th}$ row of $A$), and $f_{\emph{eff}}=2f(1-f)$.
The first $k$ symbols of $z$ are therefore a result of a
\emph{hidden} Markov Model (HMM). The underlaying transition
matrix, $T$, generating the source sequence, can be estimated by
means of the EM algorithm \cite{EM} , which is a standard tool for
solving such \emph{Parametric Estimation} problems, with linear
complexity. Having $T$ (approximately) revealed, the DBP's can be
calculated as described in Eq. (\ref{Prior(b)}).

For the general construction of the MN algorithm one
adds/subtracts rows of the concatenated matrix $[AB]$ and the
corresponding symbols in $z$, such that a situation is finally
reached as follows: The first $k$ rows of $A$ are the identity
matrix, regardless of the construction of the first $k$ rows of
$B$. From the knowledge of the noise level $f$ and the structure
of $i^{th}$ row of $B$ one can now calculate the effective noise,
$f_{\emph{i,eff}}$ , of the $i^{th}$ received source symbol. Since
all $\{f_{\emph{i,eff}}\}$ are functions of a unique noise level
$f$, one can estimate the parameters of the Markovian process
using some variants of the EM algorithm. Note, that in the general
case the first $k$ rows of $B$ contain loops, hence
$\{f_{\emph{i,eff}} \}$, are correlated. However, these
correlations are assumed to be small as the typical loop size is
of $O(\log (k))$\cite{erdos}.

\section{CONCLUDING REMARKS}

The only remaining major drawback of the presented decoder is that
the complexity (per iteration), scales as $O(kq^2)$, this may
considerably slow down the decoder even for moderate alphabet
size. Note however, that for large $q$, such that $q^2\geq k$, and
low entropy sequences, the transition matrix, $T$, is expected to
be very sparse, and dominated by elements of $O(1)$. Taking
advantage of the sparseness of $T$, the complexity of the decoder
can be further reduced.

The one-dimensional Markovian decoder can be easily extended to
coding of a two-dimensional array of symbols or even to an array
of symbols in higher dimensions \cite{Measur}. The naive
complexity of the DBP calculation scales as $k^dq^{2d+1}$, where
$k^d$ is the number of blocks in the array, and $d$ denotes the
dimension. Using Markovian and Bayesian assumptions, the
complexity can be reduced to $O(k^dq^2)$.



\begin{thebibliography}{999}
\bibitem{Tgf4} For the sake of brevity we report $T$ only for $q\!\!=\!\!4$:
\begin{equation}\label{T_GF4}
    T=\left(\begin{array}{cccc}
              0.85 & 0.07 & 0.04 & 0.04 \\
              0.23 & 0.25 & 0.05 & 0.47 \\
              0.47 & 0.05 & 0.25 & 0.23 \\
              0.04 & 0.04 & 0.07 & 0.85 \\
            \end{array}
            \right)
\end{equation}
\end{thebibliography}
\end{document}